\documentclass[twoside,a4paper,11pt]{sea10}
\usepackage{graphicx}
\usepackage{hyperref}
\usepackage{movie15}
\begin{document}
\pagenumbering{arabic}
\pagestyle{myheadings}
\thispagestyle{empty}
{\flushleft\includegraphics[width=\textwidth,bb=58 650 590 680]{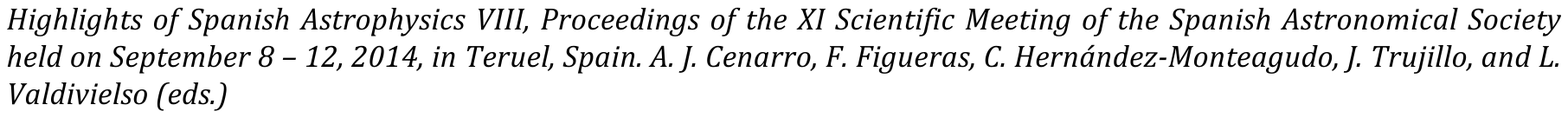}}
\vspace*{0.2cm}
\begin{flushleft}
{\bf {\LARGE
%
s-process enrichment in the planetary nebula NGC\,3918
%
}\\
\vspace*{1cm}
%
Garc\'{\i}a-Rojas, J.$^{1,2}$,
Madonna, S.$^{1,2}$,
Luridiana, V.$^{1,2}$,
Sterling, N.~C.$^{3}$,
and 
Morisset, C.$^{4}$
%
}\\
\vspace*{0.5cm}
%
$^{1}$
Instituto de Astrof\'{\i}sica de Canarias, E-38200, La Laguna, Tenerife, Spain\\
$^{2}$
Universidad de La Laguna, Departamento de Astrof\'{\i}sica,  E-38204, La Laguna, Tenerife, Spain\\
$^{3}$
Department of Physics, University of West Georgia, 1601 Maple Street, Carrolton, GA 30118, USA
$^{4}$
Instituto de Astronom\'{\i}a, Universidad Nacional Aut\'onoma de M\'exico, Apdo. Postal 70264, Mex. D. F. 04510, Mexico
%
\end{flushleft}
%
\markboth{
s-process enrichments in NGC\,3918.
}{ 
%
Garc\'{\i}a-Rojas et al. 
%
}
\thispagestyle{empty}
\vspace*{0.4cm}
\begin{minipage}[l]{0.09\textwidth}
\ 
\end{minipage}
\begin{minipage}[r]{0.9\textwidth}
\vspace{1cm}
\section*{Abstract}{\small
%
 We present deep, high-resolution (R$\sim$40000) UVES at VLT spectrophotometric data of the planetary nebula NGC\,3918. This is one of the deepest spectra ever taken of a planetary nebula. We have identified and measured more than 700 emission lines and, in particular, we have detected very faint lines of several neutron-capture elements (s-process elements:  Kr, Xe and Rb) that enable us to compute their chemical abundances with unprecedented accuracy, thus constraining the efficiency of the s-process and convective dredge-up.

%
\normalsize}
\end{minipage}
%
%
%
\section{Introduction \label{intro}}
 
About half of the heavy elements ($Z>$30) in the Universe are formed by s-process nucleosynthesis in asymptotic giant branch (AGB) stars. Several of these elements, whose detection is difficult or uncertain in AGB stars, can be detected in planetary nebulae (PNe) by means of deep, high-resolution spectroscopy.  Detection of lines from a large number of neutron-capture ions can provide strong constraints to models studying the efficiency of the s-process and convective dredge-up.

These data will also make it possible to verify the consistency of current ICFs based on photoionization modelling, which are critical to hone nebular spectroscopy into an effective tool for studying s-process nucleosynthesis.

\begin{figure}
\center
\includegraphics[scale=0.35]{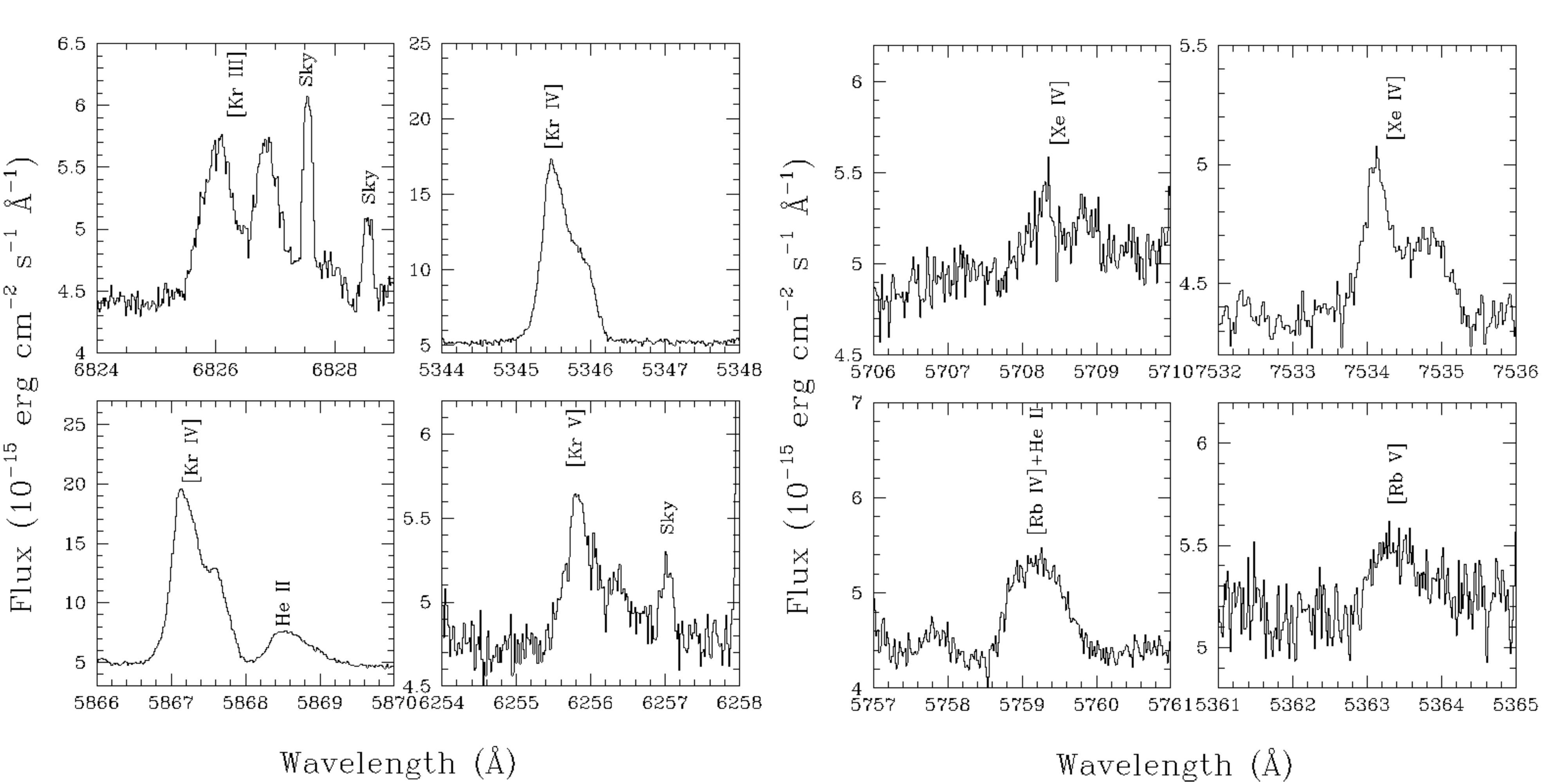}
\caption{\label{fig1} Sections of the echelle spectrum of NGC\,3928 showing some of the neutron-capture element ionic lines detected.}
\end{figure}

\section{Observations \label{obs}}

The spectra of NGC\,3918 were taken with the Ultraviolet-Visual Echelle Spectrograph (UVES, \cite{dodoricoetal00}), attached to the 8.1m Kueyen (UT2) Very Large Telescope at Cerro Paranal Observatory (Chile) in service mode on 2013 March 8.

We used the two standard settings DIC1 (346+580) and DIC2 (437+860) in both the red and blue arms of the telescope, covering the full optical range: 3100--10420 \AA\. We obtained 6 exposures of 850s each giving a total exposure time of 1.42h in each configuration. We have to emphasize that, at the end of the observing period, only 42.8\% of the total requested observing time was completed. Fortunately, the spectrum was deep enough to reach part of our scientific goals (detection of faint neutron-capture element lines). Additional single short exposures of 60 s each were taken to obtain non-saturated flux measurements for the brightest emission lines.
The slit width was set to 1$''$, which gives an effective spectral resolution of $\Delta\lambda/\lambda$$\sim$40,000 (6.5 km s$^{-1}$), which is needed to deblend some important neutron-capture emission lines from other spectral features (\cite{sharpeeetal07}). 
The slit center was set 3.8$''$ north to the central star of NGC\,3918 oriented E-W (PA=90$^\circ$) covering the brightest area of NGC\,3918. The data were reduced using the UVES-ESO pipeline to obtain a fully reduced 2D spectra. The final one-dimensional spectra we analysed were extracted for a common area for all the spectral ranges of 9.35 arcsec$^2$. The star LTT 3218 was observed to perform the flux calibration and was also fully reduced with the pipeline. 

We have identified and measured more than 700 emission lines. Important advantages of our data are the high spectral resolution and the strong  dependence of the line profile on the excitation of the line, which enable us to robustly distiguish the faint neutron-capture lines from other close-by emission lines. 

\section{Preliminary results \label{results}} 

\subsection {Physical conditions}

The large number of emission lines identified and measured in the spectrum of NGC\,3918 enables us to derive physical conditions using multiple emission-line ratios. 
The computations of physical conditions and chemical abundances were carried out with PyNeb \cite{luridianaetal14}, a python-based package
that generalizes the tasks of the IRAF package {\it nebular}. The methodology followed for the derivation of $n_e$ and $T_e$ has been described in previous papers of our group. We have updated the atomic dataset to state-of-the-art atomic data for CELs and ORLs. Errors were computed via Monte Carlo simulations.

To calculate the ionic abundances we assumed a three-zone ionization scheme. For $n_e$, we obtained very similar results from the different diagnostics used ([O~{\sc ii}], [S~{\sc ii}], [Cl~{\sc iii}] and [Ar~{\sc iv}]), so we adopted an average $n_e$=6200$\pm$1250 cm$^{-3}$  as representative for the whole nebula. For $T_e$, in the low-ionization zone we adopted the average of $T_e$([N~{\sc ii}]), $T_e$([S~{\sc ii}]) and $T_e$([O~{\sc ii}]), which amounts to $T_e$=11000$\pm$1350 K. $T_e$([N~{\sc ii}]) and $T_e$([O~{\sc ii}]) were corrected from recombination contribution to [N~{\sc ii}] $\lambda$5755 and the [O~{\sc ii}] $\lambda$7320+30 lines.
The average of $T_e$([O~{\sc iii}]), $T_e$([Ar~{\sc iii}]), $T_e$([S~{\sc iii}]) and $T_e$([Cl~{\sc iv}]), $T_e$=12150$\pm$300 K, was assumed as representative of the medium-ionization zone. Finally, $T_e$([Ar~{\sc v}])=15400$\pm$800 K was adopted as representative of the high-ionization zone.

\subsection {Chemical abundances}

Thanks to the depth of our spectrum and to the high excitation of NGC\,3918, we could compute ionic chemical abundances for a large number of ions of N$^+$, O$^+$,O$^{2+}$), Ne$^{2+}$-Ne$^{4+}$, Na$^{3+}$, S$^+$, S$^{2+}$, Cl$^+$-Cl$^{3+}$, Ar$^{2+}$-Ar$^{4+}$, K$^{3+}$, K$^{4+}$, Fe$^{2+}$-Fe$^{6+}$,  Kr$^{2+}$-Kr$^{4+}$, Rb$^{3+}$, Rb $^{4+}$), Xe ($^{2+}$, Xe$^{3+}$, and Xe$^{5+}$) from CELs. We took advantage of the large atomic data set available in PyNeb to perform these computations. We also computed ionic abundances from ORLs for some ions of C$^{2+}$-C$^{4+}$, O$^{+}$, O$^{2+}$, N$^{2+}$, Ne$^{2+}$ and Mg$^{2+}$. 

Total  abundances for N, C, O, Ne, S, Cl and Ar were computed by using the ionization corrections factors (ICFs) recently published by \cite{delgadoingladaetal14}. For Xe and Rb, the total abundance was simply computed as the sum of all the observed ionic species. This give us a lower limit to the total abundance, but we do not expect a large contribution of the unobserved ions. For Kr we adopted different ICFs: the recipes provided by \cite{sterlingetal07} and different ICFs depending on the detected ions of Kr developed by \cite{sterlingetal15}. We discarded the ICFs proposed by \cite{sharpeeetal07} for Kr and Xe because, in both cases, the total abundance was lower than the sum of the obserevd ionic species. 
We note that results obtained by using the different ICFs proposed by \cite{sterlingetal07} and \cite{sterlingetal15} display a low dispersion, suggesting that the formulae proposed are quite robust. 

\subsection{s-process enrichment in NGC~3918 \label{enrich}} 

In Table~\ref{tab1} we present the [Kr/O], [Xe/O] and [Rb/O] ratios from CELs in NGC\,3918. We use the solar abundances reported by \cite{lodders10}. We selected O as the reference element following the arguments given by \cite{sterlingdinerstein08} and based on the fact that the N/O and He/H ratios computed in NGC\,3918 suggest that it is a non-type I PN. 
In order to discern s-process enrichment in the progenitor stars from primordial scatter, \cite{sterlingdinerstein08} established that, for Galactic disk PNe with aproximately solar metallicities, s-process enrichment in excess of 0.2-0.3 dex relative to solar are the ones that can generally be attributed to s-process nucleosynthesis in their progenitor stars. Given our [s-element/O] ratios, it seems that Kr is enriched in NGC\,3918, although for Xe and Rb such enrichment is not evident. However, taking into account that our derived Xe/H and Rb/H ratios are lower limits, we can not discard that these elements are also enriched in this PN. Detailed photoionization model grids, similar to what \cite{sterlingetal07} and \cite{sterlingetal15} have done for Se and Kr, will help us to compute accurate values of the total abundances for these two elements. 
In this sense, an important advantage of our data is that we detect several ions for Kr, Xe and Rb, making these data excellent to constrain the predictions of photoionization models which, in turn, could be safely used to generate robust ICF expressions (urgently needed for Xe and Rb). At this moment, we can only conclude that the data obtained for Kr support the robustness of the particular ICFs used.

\begin{table}[ht] 
\caption{Total abundance ratios} 
\center
\begin{minipage}{0.5\textwidth}
\center
\begin{tabular}{lc} 
\hline\hline 
Ratio 	& [X/O]\footnote{[X/O]=log(X/O)-log(X/O)$_{\odot}$. Solar abundances from \cite{lodders10}.} 	\\
\hline 
[Kr/O]\footnote{ICF for Kr is an average of the ICFs proposed by \cite{sharpeeetal07}, \cite{sterlingetal07} and \cite{sterlingetal15}.}  &  0.66$\pm$0.15	\\

[Rb/O]\footnote{Lower limit. Total Rb from the sum of the ionic abundances.}   	&  0.20$\pm$0.18	\\

[Xe/O]\footnote{Lower limit. Total Xe from the sum of the ionic abundances.}   	&  0.30$\pm$0.14	\\
\hline
\end{tabular}
\end{minipage}
\label{tab1} 
\end{table}

With these data, the comparison with s-process nucleosynthesis models in AGB stars is now straightforward. We have done the exercise of comparing our results with the widely-used models by \cite{bussoetal01} for a 1.5M$_{\odot}$ progenitor star at solar metallicity. These models predict that Kr and Rb should present similar enrichment factors and that they  are larger than the enrichment exhibited by Xe. This result also reinforces our conclusion that accurate Xe and Rb ICFs are urgently needed.

\begin{figure}
\center
\includegraphics[scale=0.35]{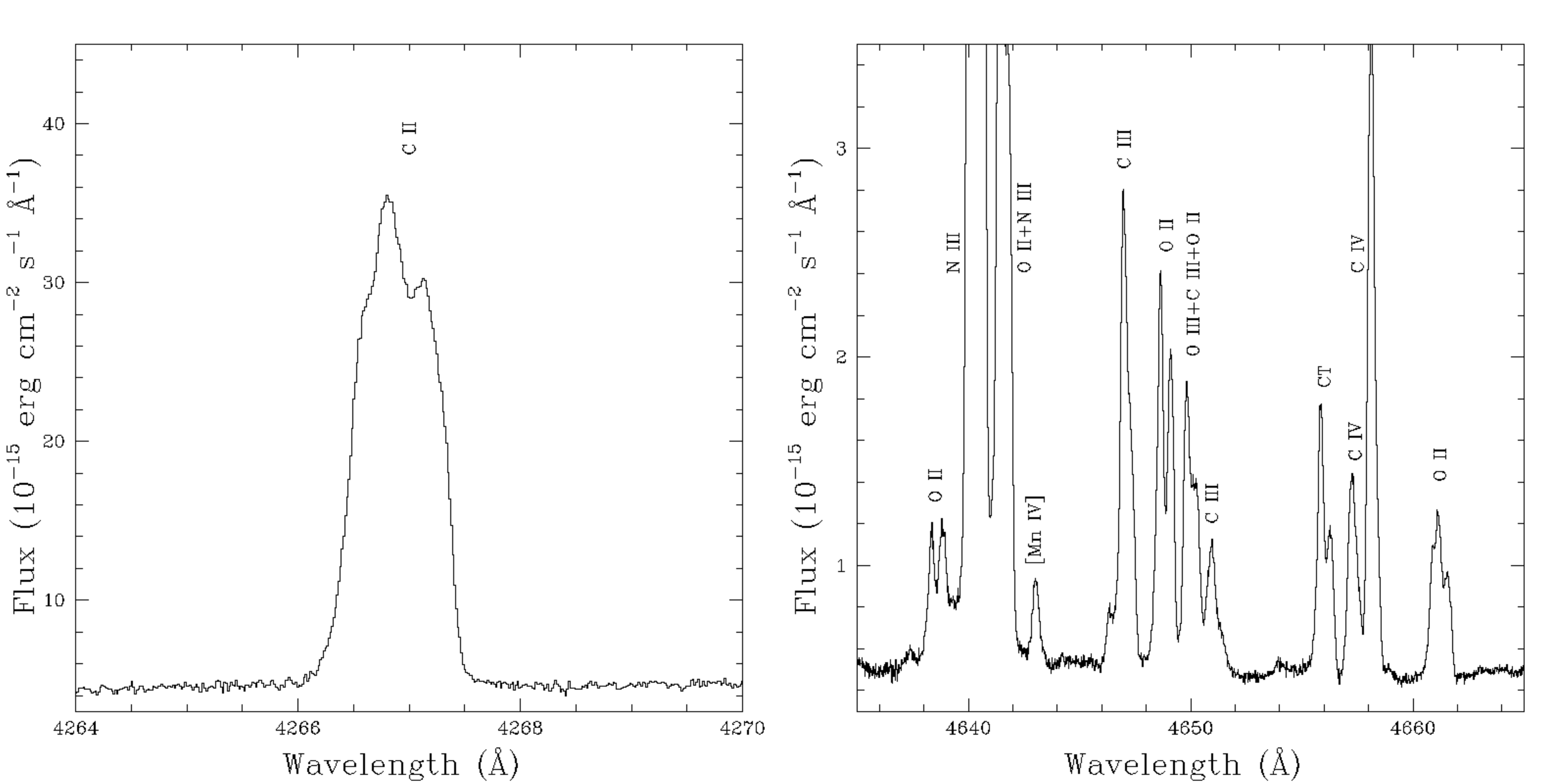}
\caption{\label{fig2} Sections of the echelle spectrum of NGC\,3918 showing the C~{\sc ii} $\lambda$4267 \AA\ recombination line and several lines of the O~{\sc ii} multiplet 1 around $\lambda$4650 \AA\. The feature marked as CT is a line produced by charge transfer from other echelle orders.}
\end{figure}

Nucleosynthesis models also predict that neutron-capture elements are expected to correlate with the C/O ratio, as carbon is brought to the surface of AGB stars along with s-processed material during third dredge-up episodes (\cite{ bussoetal01, gallinoetal98}).
In Figure~\ref{fig2} we show the correlation found by \cite{sterlingdinerstein08} for PNe of their sample with known abundances of C (from UV lines) and their Kr abundances obtained from near IR [Kr~{\sc iii}] line detections. This correlation was found to be marginal due to the large uncertainties in the Kr/H and C/O ratios. We have overplotted our result for NGC\,3819 from optical data for both C/O and Kr/O ratios (red dot; the size represent the uncertainties of the data). 
We obtained C/O=1.00$\pm$0.29 from C~{\sc ii} and O~{\sc ii} recombination lines (see~Fig~\ref{fig2} and the ICFs given by \cite{delgadoingladaetal14}. 
As it can be shown, our approach gives more precise results than the combined UV and near-IR data, and minimizes the uncertainties given by aperture effects in the different wavelength ranges as well as the uncertainties related to extinction. Additional very deep optical observations of a large sample of PNe could help to confirm this correlation of [Kr/O] with C/O and also of [Rb/O] and [Xe/O]) to constrain nucleosynthesis models.

\begin{figure}
\center
\includegraphics[scale=0.9]{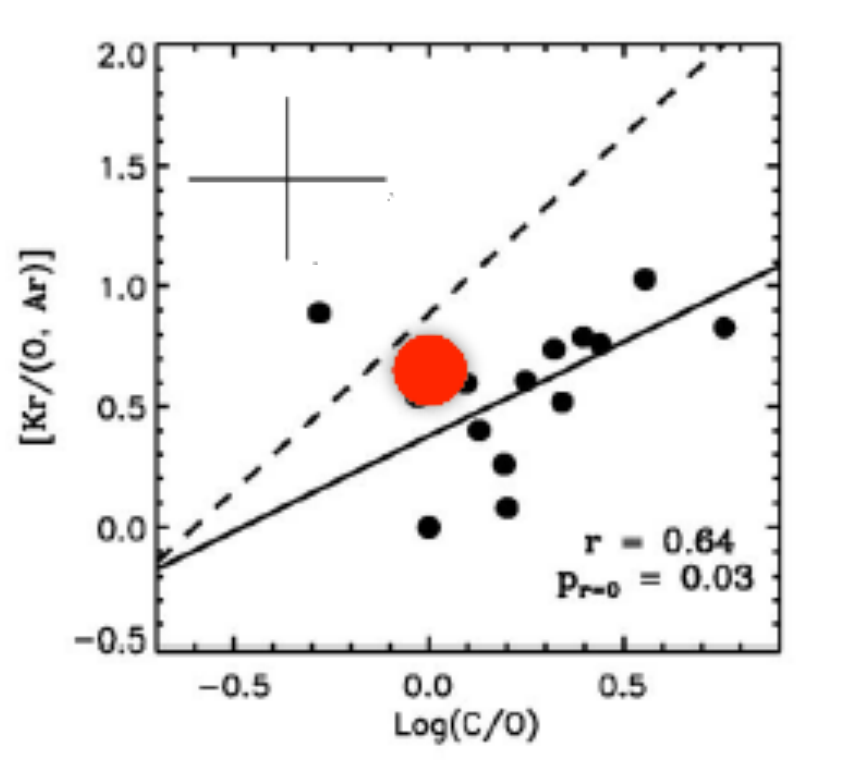}
\caption{\label{fig3} Figure 7 of \cite{sterlingdinerstein08} with the results obtained for NGC\,3918 overplotted as a red dot with semiaxis equal to the uncertainties.}
\end{figure}

A final issue to be addressed, which we can only mention here, is the comparison between 
the s-process enrichment of planetary nebulae with progenitor stars of 
different masses. From theory, we know that intermediate-mass stars experience 
smaller s-process enrichment than low-mass stars. This is due to the fact that 
the intershells of IMS are small and have little mass, so the amount of material available to the s-process nucleosynthesis is limited. Moreover, the large 
envelope masses of these objects dilute the processed material dredged up to the surface. Therefore, future works, with a larger sample of objects spanning a range in progenitor masses, will be useful to test this prediction.
%
%
\small  
%
\section*{Acknowledgments}   
%
This work has been funded by the Spanish Ministry of Economy and Competitiveness (MINECO) under the grant AYA2011-22614. JGR acknowledges support from Severo Ochoa excellence program (SEV-2011-0187) postdoctoral fellowship. This work has been developed as a MSc thesis by SM.
%

%
\end{document}